\documentclass{elsart5p}
\usepackage{graphicx}
\usepackage{amssymb}
\usepackage{url}

\begin{document}
\begin{frontmatter}

\title{Capability of Cherenkov Telescopes to Observe Ultra-fast Optical Flares}

\author{{C.~Deil\corauthref{add1}}}
\author{, W.~Domainko}
\author{, G.~Hermann}
\author{, A.-C.~Clapson}
\author{, A.~F\"orster}
\author{, C.~van~Eldik}
\author{, W.~Hofmann}

\address{Max-Planck-Institute for Nuclear Physics, P.O.~Box~103980,
  69029~Heidelberg, Germany}

\corauth[add1]{Corresponding author, Christoph.Deil@mpi-hd.mpg.de}

\begin{abstract}
The large optical reflector ($\sim 100\,$m$^2$) of a H.E.S.S. Cherenkov telescope was used to search for very fast optical transients of astrophysical origin. 43~hours of observations targeting stellar-mass black holes and neutron stars were obtained using a dedicated photometer with microsecond time resolution. The photometer consists of seven photomultiplier tube pixels: a central one to monitor the target and a surrounding ring of six pixels to veto background events. The light curves of all pixels were recorded continuously and were searched offline with a matched-filtering technique for flares with a duration of 2\,$\mu$s\,--\,100\,ms. As expected, many unresolved ($<3\,\mu$s) and many long ($>500\,\mu$s) background events originating in the earth's atmosphere were detected. In the time range 3\,--\,500\,$\mu$s the measurement is essentially background-free, with only eight events detected in 43\,h; five from lightning and three presumably from a piece of space debris. The detection of flashes of brightness $\sim 0.1$\,Jy and only 20\,$\mu$s duration from the space debris shows the potential of this setup to find rare optical flares on timescales of tens of microseconds. This timescale corresponds to the light crossing time of stellar-mass black holes and neutron stars.
\end{abstract}


\end{frontmatter}

\section{Introduction}
\label{sec:introduction}

At optical wavelengths, the sky has been surveyed in deep exposures typically lasting minutes to hours. Much faster optical flares may have escaped detection, since the dynamic sky at sub-second timescales has rarely been explored in the past \cite{eichler00}. The availability of high quantum-efficiency, fast photodetectors (photomultiplier tubes (PMTs), avalanche photodiodes or charge-coupled devices (CCDs)) has lead a few groups to construct high-time-resolution photometers, imagers or spectrographs and operate them on medium to large optical telescopes \cite{shvartsman97, straubmeier01, ryan06, dhillon07, hormuth08}.

The very large light collecting areas ($>100\,$m$^2$) of reflectors of imaging Cherenkov telescopes such as H.E.S.S. \cite{hinton04}, MAGIC \cite{lorenz04} and VERITAS \cite{krennrich04} provide an interesting alternative to normal optical telescopes \cite{eichler00}. Up to now, the only optical observations performed using the reflector of a Cherenkov telescope concerned the measurement of the optical pulsed emission from the Crab pulsar \cite{hinton06, lucarelli08}.

Compact objects like neutron stars and stellar-mass black holes are prime candidates as sources of fast optical flares. The fastest expected time variability of the emission from these systems is given by the light travel time through the immediate vicinity of the compact object, 10\,--\,100\,$\mu$s for a typical size of tens of kilometers \cite{eichler00}. Such short flares can be produced by the accretion of lumps of matter, for instance in X-ray binary systems from the stellar companion \cite{dolan01}. Relativistic outflows can further shorten the observed timescales by Doppler boosting (for exceptional examples in the case of super-massive black holes see \cite{gaidos96, aharonian07, albert07}).
Optical variability in the millisecond range has been observed for pulsars \cite{shearer03} and X-ray binaries \cite{bartolini94, beskin94, kanbach01} and has been searched for in objects known to exhibit flares in other wavelength regions, for instance at radio frequencies from rotating radio transients (RRATs) \cite{mclaughlin06, dhillon06}. No optical variability on timescales shorter than 100\,$\mu$s has been detected from any astronomical target so far.

When trying to detect astronomical flares, natural and artificial flaring events occurring in the earth's atmosphere or orbit have to be considered. Known producers of such terrestial optical flares on sub-second timescales are airplanes, satellites, lightning and shooting stars \cite{schaefer87}. Additionally, cosmic-ray induced air showers produce light flashes lasting a few nanoseconds \cite{galbraight53}.

This paper reports on 43\,h of high-time-resolution optical photometry of X-ray binaries and one RRAT using the reflector of a H.E.S.S. Cherenkov telescope during moonshine. A comparison of the performance of H.E.S.S. with regular optical telescopes for such fast photometry observations is presented in Sec.~\ref{sec:comparison}. The custom-built detector that was used to record continuous light curves at microsecond time resolution is described in Sec.~\ref{sec:camera}. An overview of the observations and a description of the flare-finding algorithm that was used to search for flares on timescales of 2\,$\mu$s\,--\,100\,ms is given in Sec.~\ref{sec:observations}. The events that were detected are presented in Sec.~\ref{sec:results}, and Sec.~\ref{sec:outlook} gives an outlook on how this new technique might be improved in the future.

\section{Comparison with Regular Optical Telescopes}
\label{sec:comparison}

The main difference in optical performance between Cherenkov and regular optical telescopes is that Cherenkov telescopes have huge reflectors but poor angular resolution, whereas optical telescopes have smaller reflectors and good angular resolution (see also \cite{beskin99}). Thus we assume here that the sensitivity of a Cherenkov telescope is limited by the shot noise from night-sky background (NSB) photons, whereas the optical telescope is considered completely background-free and photon-limited. 

Consider a flare of known duration $\tau$ occuring at a known point in time. This scenario might happen when a flare is detected in X-ray or radio observations and searched for in the optical band. For an optical telescope of diameter $D_\mathrm{O}$, the minimum detectable flux $\phi_\mathrm{O}$~[\,ph\,m$^{-2}$\,s$^{-1}$\,] is given by the requirement that it has to collect at least one photon,

\begin{equation}
\phi_\mathrm{O}\sim \frac{4}{\pi} \frac{1}{D_\mathrm{O}^2\tau}.
\label{eqn:SensitivitySignalLimited}
\end{equation}

The detection limit for a Cherenkov telescope is given by the condition that the number of photons detected from the source $S=\phi_\mathrm{C}\frac{\pi}{4}D_\mathrm{C}^2\tau$ must be at least as large as the fluctuations in the number of photons from the NSB during the flare. The photodetector aperture $\sigma_\mathrm{C}$ [rad] must be chosen to be approximately equal to the width of the point spread function (PSF) of the telescope. For an isotropic night-sky surface brightness $\psi_\mathrm{NSB}$~[\,ph\,m$^{-2}$\,s$^{-1}$sr$^{-1}$\,], telescope diameter $D_\mathrm{C}$ and flare duration $\tau$, the average number of NSB photons during the flare will be $B=\psi_\mathrm{NSB}(\frac{\pi}{4})^2D_\mathrm{C}^2\sigma_\mathrm{C}^2\tau$ and the Poisson noise $N=\sqrt{B}$. 

Solving the equation $S/N\sim 1$ for $\phi_\mathrm{C}$ gives the minimum detectable flux
\begin{equation}
\phi_\mathrm{C}\sim\frac{\sigma_\mathrm{C}}{D_\mathrm{C}}\frac{\sqrt{\psi_\mathrm{NSB}}}{\sqrt{\tau}},
\label{eqn:SensitivityBackgroundLimited}
\end{equation}
and hence
\begin{equation}
\frac{\phi_\mathrm{C}}{\phi_\mathrm{O}}\sim
\frac{\pi}{4} \frac{\sigma_\mathrm{C}}{D_\mathrm{C}} D_\mathrm{O}^2 \sqrt{\psi_\mathrm{NSB}\tau}.
\label{eqn:RelativeSensitivity}
\end{equation}

\begin{figure}
  \begin{center}
    \includegraphics[width=0.48\textwidth]{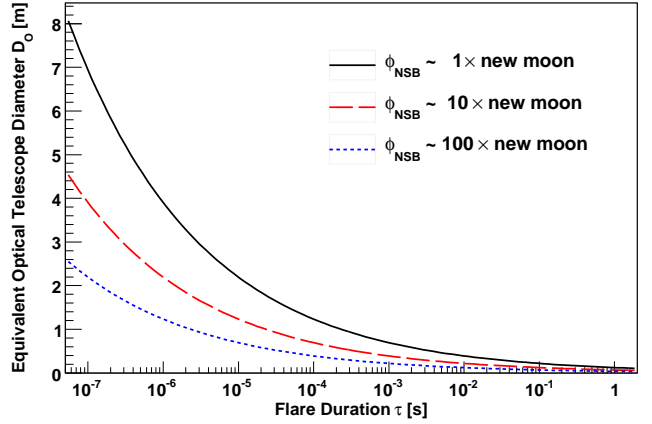}
  \end{center}
\caption{Comparison of a H.E.S.S. Cherenkov telescope (100\,m$^2$) with a normal optical telescope for high-time-resolution photometry. For a given flare duration $\tau$, the diameter of a background-free optical telescope was calculated to achieve the same sensitivity as H.E.S.S. using Eqn.~(\ref{eqn:RelativeSensitivity}) and the condition $\frac{\phi_\mathrm{C}}{\phi_\mathrm{O}}=1$ with the following parameters: H.E.S.S. telescope diameter $D_\mathrm{C}\sim 11$\,m, PSF width $\sigma_\mathrm{C}\sim 2.5$\,arcmin and typical NSB brightness without moon at the H.E.S.S. site $\psi_\mathrm{NSB}\sim 2\times 10^{12}$\,ph\,m$^{-2}$\,s$^{-1}$sr$^{-1}$ in the wavelength range 300\,--\,650\,nm \cite{preuss02}. For typical NSB levels during moonshine see Fig.~\ref{fig:moon}.}
\label{fig:OVsC}
\end{figure}

The main result from Eqn.~(\ref{eqn:RelativeSensitivity}) is that the relative sensitivity $\phi_\mathrm{C} / \phi_\mathrm{O}$ scales with $\sqrt{\tau}$, i.e. that Cherenkov telescopes perform better relative to optical telescopes for the detection of shorter flares. Inserting the values for H.E.S.S. into Eqn.~(\ref{eqn:RelativeSensitivity}), an absolute comparison with regular optical telescopes can be made. Figure~\ref{fig:OVsC} shows that even under moderate moonshine conditions Cherenkov telescopes are competitive with meter-class optical telescopes in the sub-millisecond time domain, whereas for the search of longer flares optical telescopes perform better.

For observations of high-mass X-ray binary systems with a bright companion optical telescopes will be background-limited as well. Then, the relative sensitivity of Cherenkov and optical telescopes no longer depends on the flare duration, but is given by
\begin{equation}
\frac{\phi_\mathrm{C}}{\phi_\mathrm{O}}\sim
\frac{D_\mathrm{O}}{D_\mathrm{C}}\sqrt{\frac{R_\mathrm{C,Background}}{R_\mathrm{O,Background}}},
\label{eqn:RelativeSensitivityBackground}
\end{equation}
where $R_\mathrm{Background}$ are photon detection background rates. These rates are the sum of the rates caused by light from a companion star, light from the night sky, or potentially a dark rate in the detector.

\section{The Camera System}
\label{sec:camera}

\begin{figure}
  \begin{center}
    \includegraphics[width=0.48\textwidth]{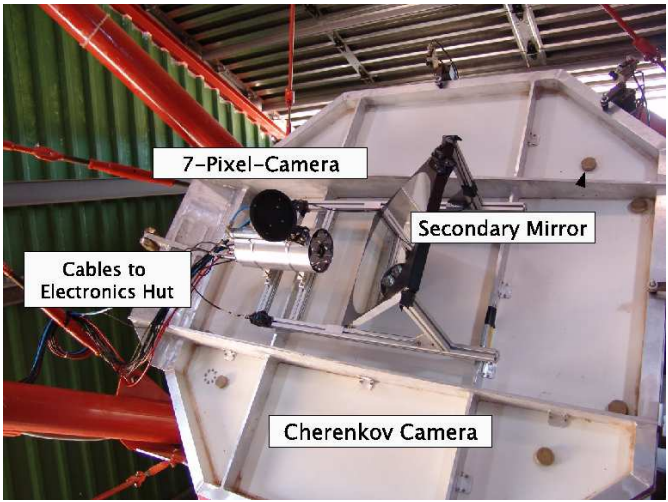}
  \end{center}
  \caption{The 7-pixel camera mounted on the lid of the H.E.S.S. Cherenkov camera.}
  \label{fig:7PixCamMounted}
\end{figure}

\begin{figure}
  \begin{center}
    \includegraphics[width=0.30\textwidth]{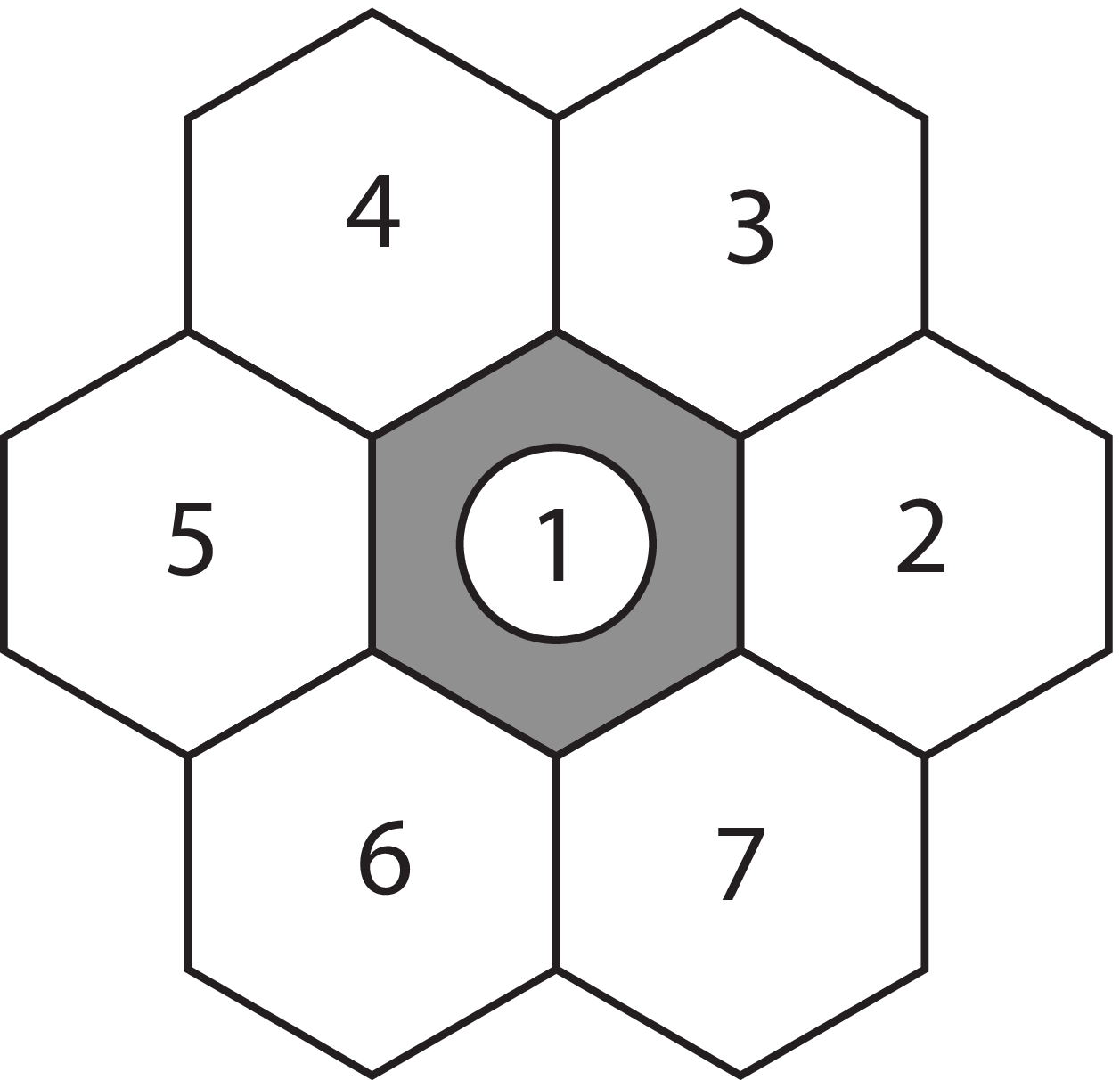}
    \includegraphics[width=0.14\textwidth]{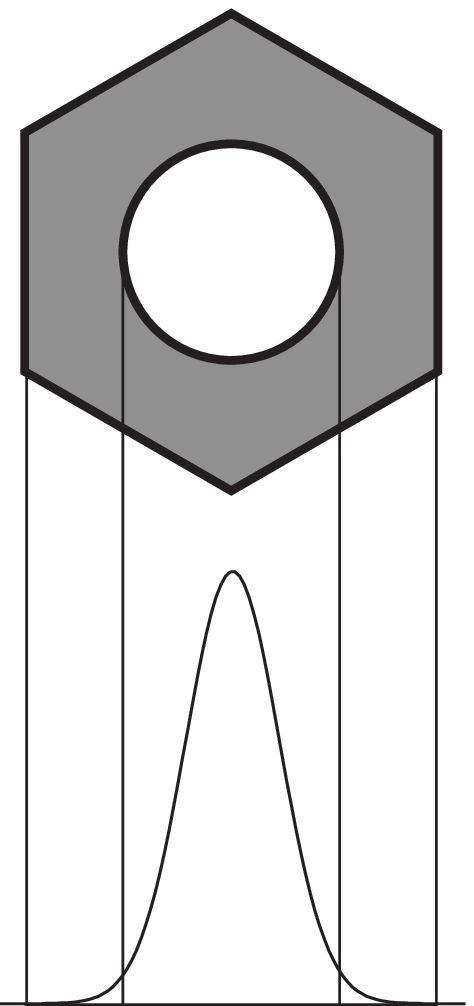}
  \end{center}
\caption{\textit{Left:} Front view of the 7-pixel camera. \textit{Right:} Comparison of the dimensions of the H.E.S.S. optical PSF full width at half maximum (FWHM$=2.5\,$arcmin), central pixel aperture diameter (5.0\,arcmin) and hexagonal outer pixel flat-to-flat width (10\,arcmin). The telescope focal length is 15\,m resulting in a plate scale of 4.4\,mm/arcmin.}
\label{fig:7PixelCamera}
\end{figure}

For the optical high-time-resolution observations, a 7-pixel camera was custom-built and mounted on the lid of the Cherenkov camera of a H.E.S.S. telescope using a plane secondary mirror to put it into focus (see Fig.~\ref{fig:7PixCamMounted}). The central pixel was used to continuously record the light curve of the astronomical target, while a ring of six `outer' pixels was used both to monitor the NSB level and as a veto system to reject terrestial background events occurring in the atmosphere (see Fig.~\ref{fig:7PixelCamera}).

For all seven pixels, Photonis PMTs of type XP2960 were used as light detectors. For this setup, the total system response mainly is in the wavelength range 350\,--\,550\,nm, roughly matching the canonical astronomical B~band (effective wavelength 445\,nm, effective width 94\,nm). In this range the average transmissivity of the atmosphere is known to be $\sim 70$\% at 15$^\circ$ zenith angle \cite{bernloehr00}. The primary and secondary mirror reflectivities (both on average $\sim 80$\%) and PMT quantum efficiency (on average $\sim 20$\%) have been measured in the lab as a function of wavelength. Assuming a flat spectrum in this range, an observed photoelectron rate can be converted into an actual flux density in Jansky.

The PMTs were equipped with hexagonal light funnels of 10\,arcmin width \cite{bernloehr03}, and the central pixel had a circular aperture of 5\,arcmin (see Fig.~\ref{fig:7PixelCamera}). This diameter was chosen such that it contains more than 90\% of the light from the target, but blocks $\sim 3/4$ of the NSB compared to the outer pixels. Note that the H.E.S.S. optical PSF has a full width at half maximum (FWHM) of 2.5\,arcmin.
The advantage of this pixel layout is that there are no gaps in the veto ring and that no part of the signal from the astronomical target will enter one of the veto pixels. However, the NSB level in the outer pixels is a factor $\sim 4$ higher compared to the central pixel, making them 2 times less sensitive to detect similar flares.

Pointing corrections for atmospheric refraction, for bending of the masts holding the Cherenkov camera, and for the offset of the central pixel relative to the optical axis of the telescope were applied. The pointing accuracy was checked frequently by observing a bright star near the target. This was done by closing the pneumatic lid of the 7-pixel camera and taking a picture of the image of the star on that lid with a CCD located between mirror segments in the center of the H.E.S.S. reflector. The remaining mispointing was below 0.5\,arcmin, compared to 5~arcmin diameter of the central pixel, resulting only in a negligible signal loss.

The analog signals from the PMTs were transmitted via 50\,m coaxial cables to an electronics container. There, they were amplified and passed through an analog low-pass filter with a cutoff near 0.5\,$\mu$s. Then, the signals were continuously digitized with a sampling period of 10\,ns, and 256 consecutive samples were averaged to reduce the data rate to 6 megabyte/s, resulting in a 2.56\,$\mu$s sampling. The absolute time of each sample is known with a precision of $\sim1\,\mu$s from a GPS clock. An additional coaxial cable was connected to a termination resistor in the 7-pixel camera and its signal was used to monitor possible electronic interference (for details see \cite{deil08}).

\section{Observations and Data Analysis}
\label{sec:observations}

\begin{figure}
  \begin{center}
    \includegraphics[width=0.48\textwidth]{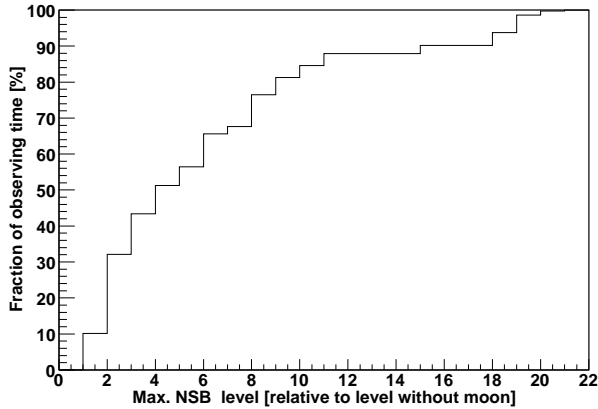}
  \end{center}
  \caption{NSB levels during the 43 hours of good quality observations in May 2\,--\,29, 2007, roughly one moon period. This integral plot shows the fraction of the observing time the NSB was below a given level, relative to the NSB level without moon ($\psi_\mathrm{NSB}\sim 2\times 10^{12}$\,ph\,m$^{-2}$\,s$^{-1}$sr$^{-1}$ in the wavelength range 300\,--\,650\,nm \cite{preuss02}). Note that about 50\% of the observations occured at NSB levels less than five times the level without moon.}
  \label{fig:moon}
\end{figure}

In May 2007 high-time-resolution observations of several X-ray binaries containing a black hole or a neutron star ({Sco~X-1}, {GX~339-4}, {MXB~1735-44}, V4641~Sgr and SS~433) and of the isolated neutron star {RRAT~J1317-5759} were performed during moonshine. After removing runs where clouds entered the field of view, 43\,h ($\sim 1$~terabyte) of good quality data remain.
The level of the NSB depends mainly on the brightness of the moon, the altitude of the target and the moon-target angular separation. Our observations range from new moon to more than $3/4$~moon, altitudes of 30$^\circ$\,--\,90$^\circ$ and moon-target separations of 30$^\circ$\,--\,120$^\circ$. The measured NSB level increases up to a factor of $\sim 20$ relative to no moon (see Fig.~\ref{fig:moon}). Note that the shot noise from the NSB, which determines the sensitivity for detecting flares, varies by a factor of $\sim\sqrt{20}$ for these conditions.

\begin{figure}
\includegraphics[width=0.48\textwidth]{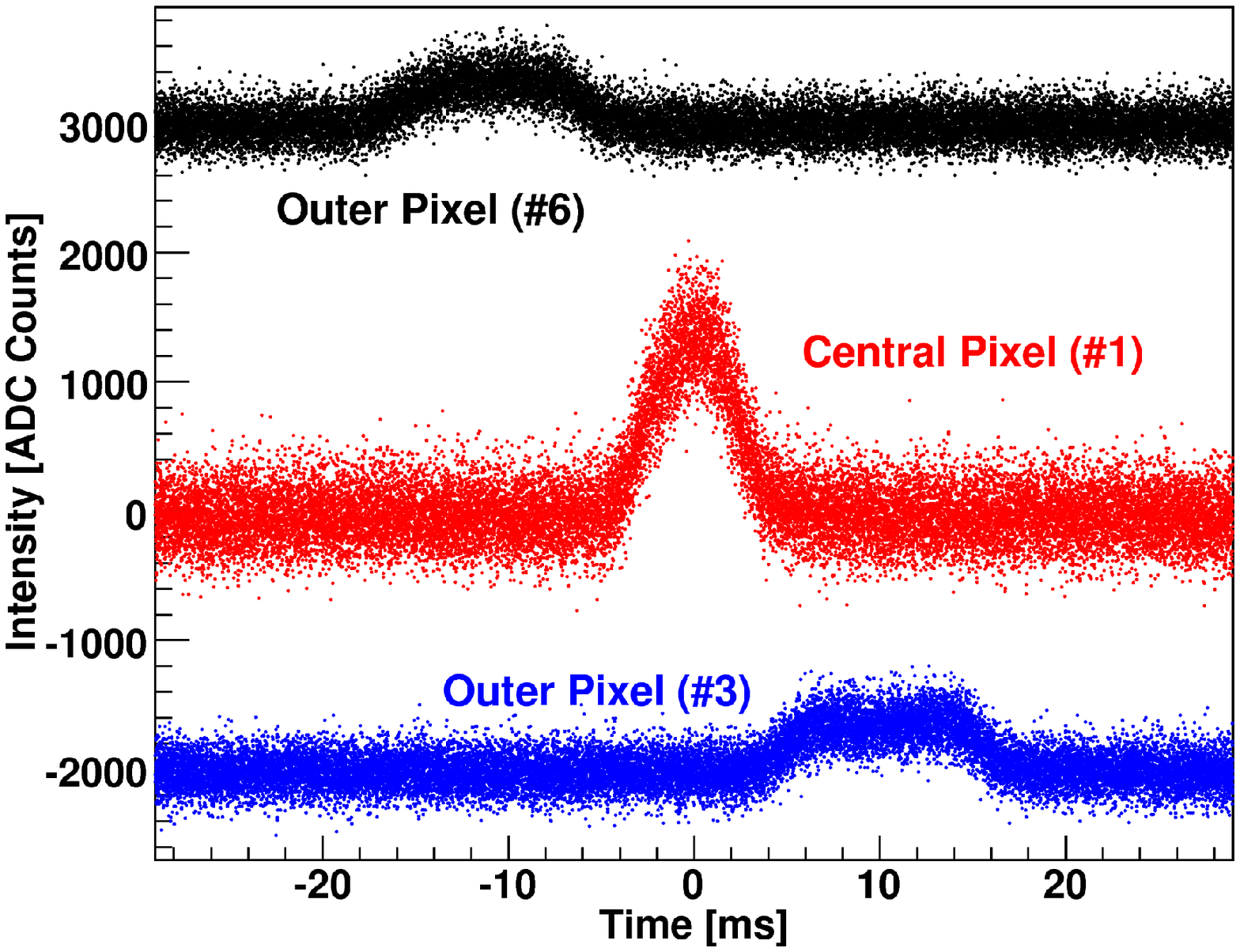}
\includegraphics[width=0.48\textwidth]{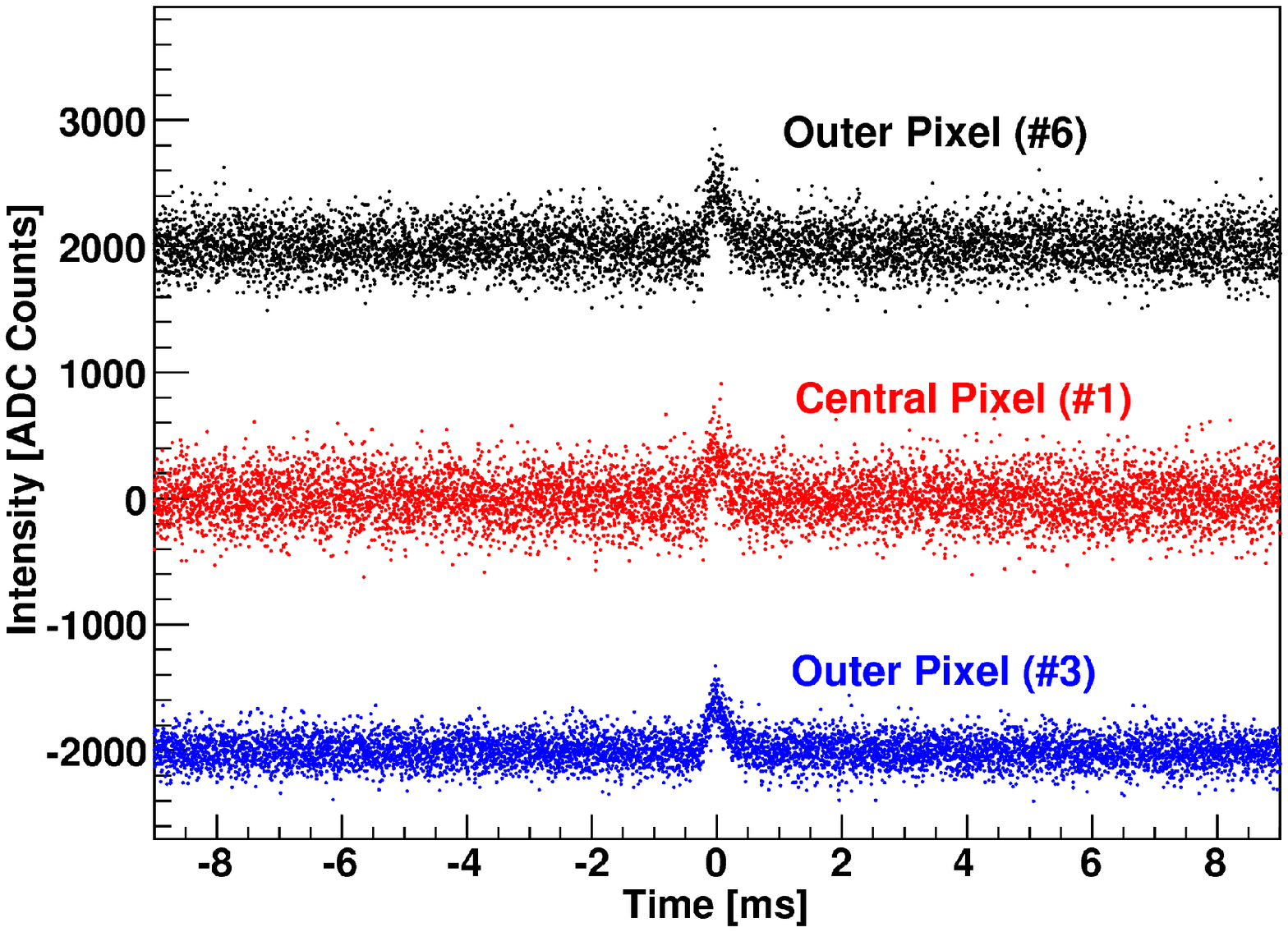}
\caption{Example background events very likely caused by a shooting star (\textit{top}) and lightning (\textit{bottom}). The intensities were arbitrarily offset such that the light curves from different channels do not overlap. }
\label{fig:BackgroundEvents}
\end{figure}

Non-astronomical events are expected to occur, producing a background of events that has to be rejected to find flares from the astronomical target. Airplanes, satellites and shooting stars  \cite{hinton06} passing through the field of view of the whole camera and also the central pixel are expected to produce time-shifted flares with the following pattern: first in one or two of the outer pixels, then in the central pixel and then in one or two of the outer pixels on the opposite side (see Fig.~\ref{fig:BackgroundEvents} top). Lightning at the horizon being scattered in the atmosphere will illuminate all pixels in the same way, making it easy to veto (see Fig.~\ref{fig:BackgroundEvents} bottom). Cosmic-ray induced air showers typically show an elliptical light distribution on the sky with diameters of $\sim\,30\,$arcmin  and last for a few nanoseconds. Such flashes will register as temporally unresolved events in our detector. Given the $\sim 30\,$arcmin field of view of the 7-pixel camera, cosmic rays seen in the central pixel should be seen simultaneously by at least one veto pixel.

The data from all seven pixels and the electronics interference monitoring channel were searched for events that stand out above the noise. The challenge of finding weak signals in a noise-dominated time series is similar to the one faced by gravitational wave detectors, and a large set of literature exists on the topic (see \cite{Jaranowski} and references therein). The complexity of the methods and their performances depend to a large degree on the assumptions that can be made both on the noise in the series and the characteristics of the signal.

Of particular interest for its simplicity is the \emph{matched filter}, where a particular waveform is assumed for the signal. A generic flare search algorithm called \emph{Peak Correlator (PC)} was derived from this principle (presented in \cite{Arnaud}, implemented in \cite{BuL}), using a bank of Gaussian templates $\exp(-t^2/2\tau^2)$ as assumed signal shapes. Gaussians roughly fit any structureless flare, and using several widths $\tau_\mathrm{i}$ makes it possible to cover a range $[\tau_\mathrm{Min},\tau_\mathrm{Max}]$ of timescales. The template spacing is chosen such that the loss of detection significance for an actual Gaussian shaped flare does not decrease by more than some $\epsilon$ for any width $\tau$ in the specified range.

The PC algorithm calculates the statistical significance of flares, assuming a Gaussian waveform, by taking the noise in the series into account, estimated by the power spectral density (PSD) of the data. In practice, the time series is split into chunks of length $T_\mathrm{Chunk}$ and each chunk into a number of windows $N_\mathrm{Windows}$ such that the length of a window $T_\mathrm{Window}=T_\mathrm{Chunk}/N_\mathrm{Window}$ is large compared to $\tau_\mathrm{Max}$. The PSD of a chunk is the average of the PSDs of its windows. The underlying assumption is that the noise level is stable on timescales longer than $T_\mathrm{Chunk}$.

\begin{figure}[tbh]
  \begin{center}
    \includegraphics[width=0.48\textwidth]{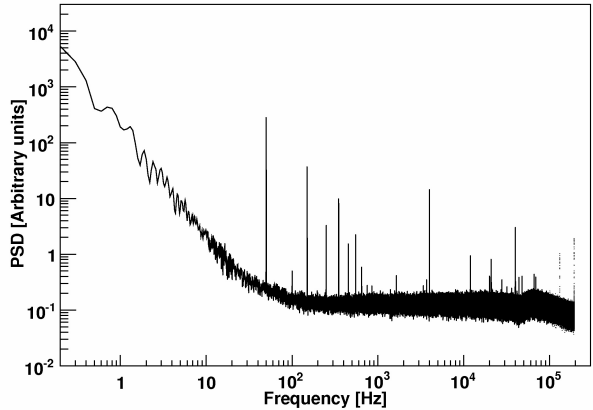}
  \end{center}
  \caption{A typical PSD calculated from one chunk of data containing no detected events (see text for explanations).}
  \label{fig:PSD}
\end{figure}

As can be seen in Fig.~\ref{fig:PSD}, above $\sim100\,$Hz (corresponding to timescales shorter than $\sim10\,$ms), the noise is approximately flat, which is characteristic of shot noise from random photon arrival times. This background is usually dominated by the NSB and sometimes by a bright star in the field of view. It changes slowly with zenith angle on timescales of many minutes while the telescope tracks the target. Additionally, due to field rotation (the H.E.S.S. telescope has an altitude-azimuth mount), a bright star sometimes rotates around the center and moves from one outer pixel to the next, changing their background levels on timescales of many minutes.
Below 100\,Hz (corresponding to timescales above 10\,ms) the signal is dominated by electronic noise ($1/f$ noise). Its level changes mainly with temperature, again on timescales much longer than $T_\mathrm{Chunk}$.
The PSD contains strong lines at 50\,Hz and multiples up to $\sim 1\,$kHz, resulting from electrical interference from the power supply. The PC algorithm correctly takes these noise components into account and prevents false event detections on the corresponding timescales.

The output of the PC filter for a given template width $\tau_\mathrm{i}$ is a time series of signal-to-noise ratio (SNR) values at the same sampling as the original light curve ($t_\mathrm{s}=2.56\,\mu$s). For an input time series of Gaussian-distributed noise, the output is also Gaussian. Thus a simple amplitude threshold $A_\mathrm{Thr,Det}$ on the SNR series can be used to trigger a detection in the central pixel, and $A_\mathrm{Thr,Veto}$ to trigger a detection in one of the six outer pixels or in the electronics interference monitoring channel. For actual parameter values used in this analysis see Tab.~\ref{tab:SearchParameters}.

\begin{table}[tbh]
\begin{center}
\small
\caption{Flare search parameters for the PC and veto algorithm (see text for explanations).}
\label{tab:SearchParameters}
\begin{tabular}{lll}
  \hline
  Parameter \hspace{2.4cm} & Symbol \hspace{1.9cm} & Value \\
  \hline
Timescale range & $[\tau_\mathrm{Min},\tau_\mathrm{Max}]$ & $[2\,\mu$s$,100\,$ms$]$ \\
Maximum significance loss & $\epsilon$ & 1\% \\
Chunk duration & $T_\mathrm{Chunk}$ & 21\,s \\
Windows per chunk & $N_\mathrm{Windows}$ & 16 \\
Detection SNR threshold & $A_\mathrm{Thr,Det}$ & 8 \\
Veto SNR threshold & $A_\mathrm{Thr,Veto}$ & 7 \\
Clustering factor & $n$ & 4 \\
Veto window & $\Delta T_\mathrm{Veto}$ & 100\,ms\\
  \hline
\end{tabular}
\end{center}
\end{table}

When consecutive samples are above threshold, the first and last ones define the start $t_\mathrm{Start}$ and stop time $t_\mathrm{Stop}$ of a so called micro-event. The SNR of the micro-event is given by the maximum SNR value in this time range, and the micro-event time is determined by the time of this maximum SNR.
A flare will trigger on several nearby timescales $\tau_\mathrm{i}$ and might contain substructures, leading to several close-by micro-events. Micro-events are clustered if their clustering windows $[t_\mathrm{Start}-n\tau_\mathrm{i}, t_\mathrm{Stop}+n\tau_\mathrm{i}]$ overlap, where $n$ is the clustering factor of Tab.~\ref{tab:SearchParameters}. The time, timescale and SNR of the clustered event is defined by the micro-event with the highest SNR. 

The following simple coincidence veto algorithm was applied to every event detected in the central pixel: if any of the six outer pixels had detected an event on any of the tested timescales within $\Delta T_\mathrm{Veto}$ of the central pixel event time, the flare was vetoed. The longest observed time shifts are 50\,ms for some shooting stars, thus a veto time window $\Delta T_\mathrm{Veto}=100\,$ms was chosen conservatively. The veto threshold $A_\mathrm{Thr,Veto}=7$ was chosen such that even the shortest timescales do not trigger on statistical fluctuations due to the large number of trial factors ($\sim 10^{11}$). Electrical interference will also influence the signal from the PMTs, although with a much lower significance, since their time series is dominated by photon shot noise, which is absent from the resistor channel. Whenever the electronics interference monitoring channel detected an interference signal within $\Delta T_\mathrm{Veto}$ of the central pixel event time, this event was rejected. In Sec.~\ref{sec:results} only events where no electronic interference occured are discussed. When applying this veto algorithm, a total of 4\% of the observation time is vetoed.

\section{Results}
\label{sec:results}

\begin{figure}
  \begin{center}
    \includegraphics[width=0.48\textwidth]{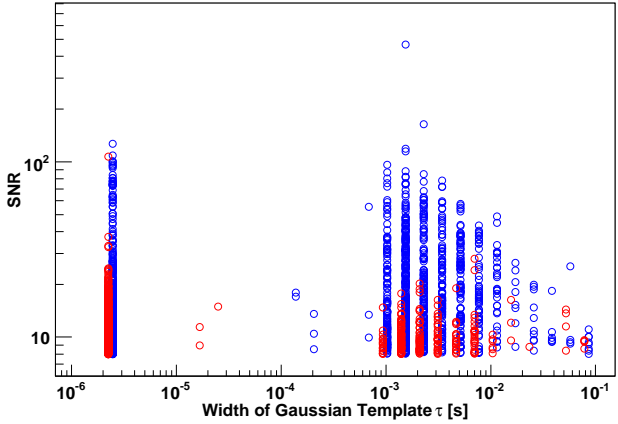}
  \end{center}
  \caption{SNR and timescale $\tau$ for all events in 43\,h of observations. Every circle represents a flare detected in the central pixel. Red events were not vetoed, blue events were. Note that blue events were shifted to the right by multiplying their $\tau$ by a factor 1.1 such that they can be seen more clearly. Events vetoed by the electronics interference monitoring channel are not shown at all.}
  \label{fig:SNR_MaxTau}
\end{figure}

Applying this flare search and veto algorithm to the data, with the parameters shown in Tab.~\ref{tab:SearchParameters}, we find the events shown in Fig.~\ref{fig:SNR_MaxTau}. On timescales above 500\,$\mu$s, a large number of 1175 events was detected. Most of these events are probably background from shooting stars, since a visual inspection of their light curves shows that most have a temporal pattern similar to the pattern of the event shown in Fig.~\ref{fig:BackgroundEvents} top. 21\% of these events are not vetoed, especially those of low SNR. This is understandable as some shooting stars might travel close to head-on towards the telescope. These events may remain in an outer pixel for some time and give a flat-topped flare that is not very Gaussian-like. 

Events with $\tau < 3\,\mu$s (1538 events, 51\% not vetoed) must be considered as temporally unresolved, since they last only one or two samples, which would also be the case for much shorter flares because of the analog shaping of 0.5\,$\mu$s. These events might be due to air showers of a few nanoseconds duration. In that case many low-SNR events might not be detected by the outer pixels because air showers exhibit large intensity variations within the shower on scales of few arcminutes. 

In the time range 3\,--\,500\,$\mu$s, where the fastest flares from stellar-mass black holes and neutron stars are expected to occur, there is considerably less backgound: only eight events were detected. The five vetoed events at $\tau\sim 120\,\mu$s are isolated events that occured in the nights of May 11 and 12. They show the same structureless waveform in all pixels and are most likely caused by single lightning flashes (Fig.~\ref{fig:BackgroundEvents} bottom shows the one with the highest SNR as an example).

\begin{figure*}
  \begin{center}
    \includegraphics[width=\textwidth]{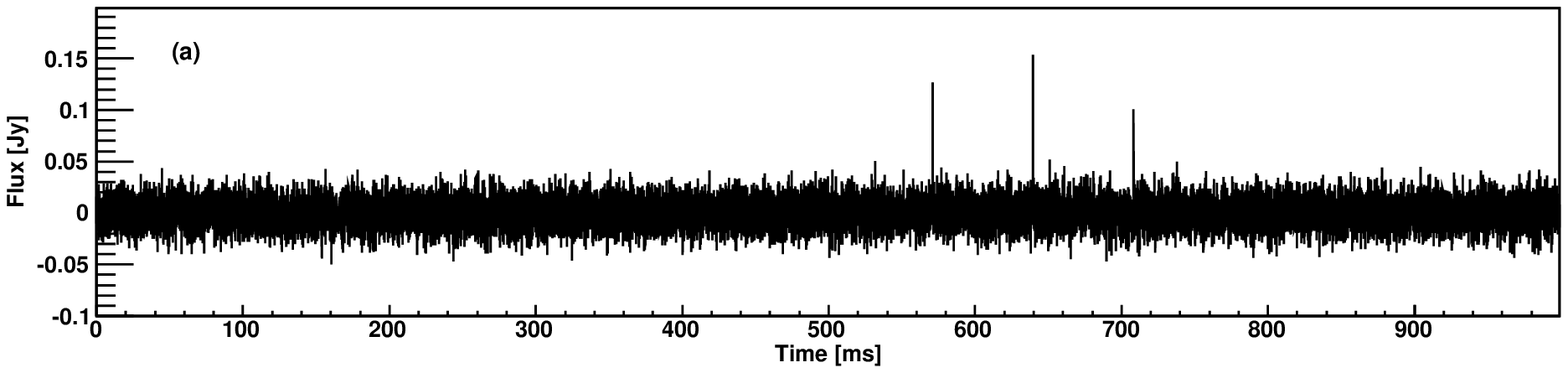}
    \hspace{1mm}
    \includegraphics[width=\textwidth]{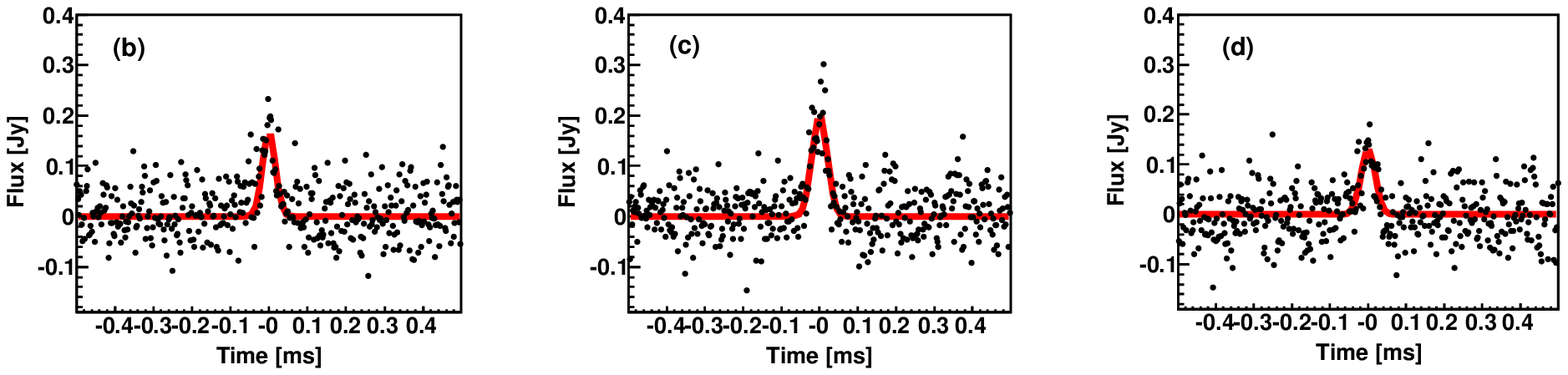}
  \end{center}
  \caption{Light curve of the three optical flares from the direction of V4641~Sgr on May~14,~2007. Panel (a) shows one second of the raw light curve with a binning of 38.4~$\mu$s (15~samples). The time is given relative to 3:47:28~UTC. (Due to technical problems with the  time synchronization of the measurement computer, the absolute time of the event might be offset by at most 10 seconds. For the relative timing between samples, the GPS signal was used directly by the analog-to-digital converter and is thus correct.)
    Panels (b), (c) and (d) each show a zoom-in of one millisecond of observations at the original binning of 2.56~$\mu$s, with the time axis centered on the flare peaks. The red curve is a fit of a Gaussian (width and amplitude as free parameters).}
  \label{fig:OpticalLightcurve}
\end{figure*}

\begin{figure}
  \begin{center}
    \includegraphics[width=0.48\textwidth]{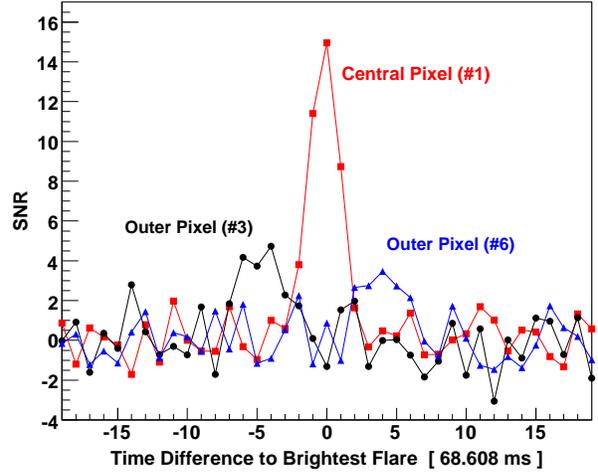}
  \end{center}
\caption{Temporal signature of the space debris events. This plot shows the SNR (for $\tau=22\;\mu$s) at multiples of $68.608\;$ms before and after the SNR\,$=15$ event in the central pixel.}
\label{fig:SpaceDebris}
\end{figure}

The three events at $\tau\sim 20\,\mu$s (light curves shown in Fig.~\ref{fig:OpticalLightcurve}) were not vetoed by any of the outer pixels. They are separated by 68.608\,(3)\,ms from each other and thus only form one exceptional event. Using the method explained in Sec.~\ref{sec:camera}, the peak flux of the brightest flare (SNR\,$= 15$ at $\tau=22\mu$s) was determined as 0.2\,Jy. The flares were recorded while observing the X-ray binary black hole candidate V4641~Sgr on May~14,~2007 at 3:47~UTC. Observing conditions were very good (altitude 63$^\circ$, 2.6~days before new moon, moon-target separation 101$^\circ$).
A careful analysis of the data shows, however, that the flares were not caused by the astronomical target, but very likely by a piece of space debris crossing the field of view. A flare separation of 68.6\,ms corresponds to a rotation frequency of 14.6\,Hz. One shiny surface would produce a beam of reflected sunlight of diameter 0.5$^\circ$, the diameter of the sun on the sky. A telescope placed in this rotating beam would see flares of approximately rectangular shape and $0.5^\circ/360^\circ \times 68.6\,$ms$\,=95\,\mu$s duration. Given that the telescope might not be located in the center of the $0.5^\circ$ circular beam but towards the edge, any flare duration shorter than but of the same order as  $95\,\mu$s is possible, in rough agreement with the observed $\tau=20\,\mu$s (corresponding flare FWHM\,$=2.4\tau=50\,\mu$s).
One consequence of this interpretation would be that at a separation of a few times 68.6\,ms further flares should be present in one of the outer pixels when the debris passed through their field of view. As can be seen in Fig.~\ref{fig:SpaceDebris}, this is actually the case. Three events at SNR\,$\sim 4$ occurred in outer pixel \#3 right before the events in the central pixel, and five events at SNR\,$\sim 3$ right after in the opposite outer pixel \#6. This shows that indeed an object was crossing the field of view of the whole camera and that the flares cannot be of astrophysical origin. Assuming a height of $\sim 1000\,$km (the object had to be above 600\,km to be outside the earth's shadow and almost all space debris is at a height of less that 2000\,km), the angular speed of 0.6$^\circ/$s corresponds to a speed of 10\,km/s, a typical value for space debris orbits \cite{UN99, NASA08}.
An airplane with strobe lights could in principle produce the observed signal, but neither the 15\,Hz repetition rate, nor the 20\,$\mu$s duration are used in airplane strobes, and the Namibian air control confirmed that there were no airplanes within 100\,km of the H.E.S.S. site at the time the flares occurred. Satellites do not have strobe lights and do not rotate as fast as 15\,Hz and can thus also be excluded as the origin of the flares.

\section{Outlook}
\label{sec:outlook}

In this paper it is demonstrated that the huge optical reflector of Cherenkov telescopes can be used to observe faint optical flashes in the sub-millisecond time domain. Flares with a duration of $\sim 20\,\mu$s as faint as 0.1\,Jy have been detected with high statistical significance using a H.E.S.S. telescope equipped with a dedicated detector during moderate moonshine.

Compact objects like neutron stars and black holes are known to produce bright and short optical flares (e.g. \cite{uemura04}), indicating a spatially small emission region. The small size of the emission region also constrains the nature of the radiation mechanism \cite{shvartsman89}: ultra-fast flashes are likely of non-thermal origin, e.g. cyclotron or synchrotron radiation \cite{kanbach01, uemura04, fabian82}. 
If the emission is generated in a relativistic jet, the shortest observed variability timescale can place a lower limit on the Doppler factor of this outflow \cite{eichler00}.

X-ray binaries have been pointed out as promising targets for sub-millisecond time resolution. Several interesting aspects related to the compact objects in these systems could be investigated:
\begin{itemize}
\item The distinction whether the compact object has a solid surface or is a black hole: a blob of matter spiraling into the event horizon of a black hole would produce a series of progressively weaker flares, whereas accretion onto a neutron star would end in a final bright flash when the matter hits its surface \cite{dolan01, stoeger80}.
\item Measurement of the black hole spin: Fukumura \& Kanzanas \cite{fukumura08} propose that photons emitted inside the ergosphere can orbit a rapidly rotating black hole. After leaving the vicinity of the black hole, they may arrive at a distant observer as double or triple flares, separated by constant time lags, which is connected to the spin of the black hole.
\item Study of the characteristics of super-Eddington flows: Another exotic possibility to create fast flashes of radiation are bubbles filled with photons, which could be created during accretion onto a neutron star \cite{klein96, jernigan00}.
\end{itemize}

For the study of bright objects, the night-sky background faced by optical observations with a Cherenkov telescope is less critical and a large reflector is especially important. In this case Cherenkov telescopes could be used to monitor such systems, since e.g. H.E.S.S. is not operated during moonshine. Significant observing time ($\sim 40\,$h$/$month) is available without interfering with normal gamma-ray observations, whereas time on large optical telescopes is expensive.

The observations presented here show that a considerable background of optical flashes with origin likely in the earth's atmosphere is present on timescales below 3\,$\mu$s and above 0.5\,ms. A ring of veto pixels has been shown to be effective, but not perfect to reject these events. A more efficient veto system could consist of a second telescope at a distance of several hundred kilometers observing the same target at the same time. Any flare from an astronomical target must occur coincident with the same waveform in both telescopes. Light sources at smaller distances will be seen by only one system if the parallax of the light emitting object is larger than the angular diameter of the central pixel. A separation of the two telescopes of 700\,km with an angular diameter of the central pixel of 0.1$^\circ$ will place any coincident event at a distance of at least 400 000\,km and thus outside the earth--moon system.

The H.E.S.S. array will be extended with a fifth, much larger telescope. This H.E.S.S. II telescope will feature a light collecting area of 600\,m$^2$, allowing more sensitive searches for ultra-fast optical transients. Continuing high-time-resolution optical photometry observations with an improved setup (higher quantum efficiency, smaller pixels, faster readout) in conjunction with a second telescope (optical or Cherenkov) at a large distance seems promising.

\appendix

\section*{Acknowledgments}

The authors would like to thank the H.E.S.S. collaboration for permission to use a H.E.S.S. telescope for these observations. We thank Thomas Kihm for help with the design and implementation of the data acquisition and Stefan Schmidt and Christian Neureuther for constructing the 7-pixel camera. Toni Hanke and Eben Tjingaete were very helpful with the installation and operation of the instrument in Namibia.

\end{document}